\documentclass[english]{article}
\usepackage[T1]{fontenc}
\usepackage[latin9]{inputenc}
\usepackage{float}
\usepackage{amstext}
\usepackage{amssymb}
\usepackage{graphicx}
\usepackage{esint}

\makeatletter

\pdfoutput=1
\usepackage{amsfonts,amssymb}
\usepackage{latexsym}
\usepackage{epsfig}
\usepackage{cite}
\usepackage{graphicx}
\usepackage[colorlinks,linkcolor=blue]{hyperref}
\newcommand{\be}{\begin{equation}}
\newcommand{\ee}{\end{equation}}

\makeatother

\usepackage{babel}
\begin{document}
{}~ \hfill\vbox{\hbox{hep-th/yymmnnn}}\break
\vskip 3.0cm
\centerline{\Large \bf Two-dimensional regular string black hole in different gauges}

\vspace*{10.0ex}
\centerline{\large Shuxuan Ying}
\vspace*{7.0ex}
\vspace*{4.0ex}
\centerline{\large \it Department of Physics, Chongqing University}
\vspace*{1.0ex}
\centerline{\large \it Chongqing, 401331, China} \vspace*{1.0ex}
\vspace*{4.0ex}

\centerline{ysxuan@cqu.edu.cn}
\vspace*{10.0ex}
\centerline{\bf Abstract} \bigskip \smallskip

This paper serves as an extended version of our previous letter arXiv:2212.03808 [hep-th]. In this paper, we  investigate the  non-perturbative and non-singular black hole solutions  derived from complete $\alpha^{\prime}$ corrected Hohm-Zwiebach action in different gauges (coordinate systems). In addition to the results that we obtained in the previous letter, we present the following additional results in this paper: 1) The regular black hole solutions in different gauges can be mutually transformed. 2) The $\alpha^{\prime}$ corrections  do not introduce any extra singularities beyond the event horizon. 3) The event horizon remains unaffected by the $\alpha^{\prime}$ corrections. The results of this paper provide valuable illustrations for further investigation of more complicated regular black hole solutions utilizing the  complete $\alpha^{\prime}$ in the near future.

\vfill \eject
\baselineskip=16pt
\vspace*{10.0ex}
\tableofcontents

\section{Introduction}

Spacetime singularities are a fundamental challenge in black hole
physics, they manifest divergent Kretschmann scalar and incomplete
geodesics at positions. In Einstein's gravity, black hole singularities
are inevitable and cannot be eliminated \cite{Penrose:1964wq,Hawking:1965mf}.
However, as Einstein's gravity is only the perturbative leading order
of a comprehensive theory of quantum gravity, it is reasonable to
assume that its solutions only describe the low-curvature regime of
spacetime. String theory, one candidate of quantum gravity, is expected
to incorporate non-perturbative effects to resolve spacetime singularities.
Approaching to spacetime singularities, string theory receives higher-curvature
corrections controlled by the squared string length $\alpha^{\prime}$.
Therefore, understanding the complete $\alpha^{\prime}$ corrections
of string theory becomes crucial step to cure the black hole singularities.

In recent works \cite{Hohm:2015doa,Hohm:2019ccp,Hohm:2019jgu}, Hohm
and Zwiebach classified all $\alpha^{\prime}$ corrections of closed
string at all orders and derived the full action using $O\left(d,d\right)$
symmetry. The reason that full action can be obtained by $O\left(d,d\right)$
symmetry lies in the fact that when closed string configurations are
independent of $m$ coordinates, the action exhibits an $O\left(m,m\right)$
symmetry, as proved by Sen in closed string field theory \cite{Sen:1991zi,Sen:1991cn}.
This observation has been verified in the tree-level string effective
action and its first-order $\alpha^{\prime}$ correction. It implies
that the low energy effective action, incorporating $\alpha^{\prime}$
correction, can always be expressed using the standard $O\left(d,d\right)$
matrix in terms of $\alpha^{\prime}$ corrected fields for the time-dependent
background \cite{Veneziano:1991ek, Sen:1991zi,Sen:1991cn, Meissner:1991zj,Meissner:1996sa}.
Based on this finding, Hohm and Zwiebach assumed that this result
holds true for all orders in $\alpha^{\prime}$. With this assumption,
Hohm and Zwiebach demonstrated that all orders $\alpha^{\prime}$
corrections can be classified according to even powers of the Hubble
parameter in the context of the FLRW cosmological background. The
dilaton field only includes first-order time derivatives. These features
enable the equations of motion (EOM) to involve solely two derivatives
of the metric and to be exactly solvable. This significant advancement
introduces a new action for studying the black holes and cosmology
within non-perturbative string effects. By using this action, numerous
new regular black hole solutions and cosmological solutions had been
studied. In refs. \cite{Wang:2019kez,Wang:2019dcj,Wang:2019mwi,Gasperini:2023tus,Song:2023txa},
the cosmological big bang singularity have been successfully removed.
Moreover, the curvature singularities of two-dimensional string black
hole are also resolved in refs. \cite{Ying:2022xaj,Ying:2022cix,Codina:2023fhy}.

This paper is an extended version of our previous letter \cite{Ying:2022xaj}.
In this paper, we investigate the two-dimensional regular black hole
solution derived from the Hohm-Zwiebach action using two distinct
ansatz. Firstly, we consider the ``Schwarzschild'' gauge, which
relies on the radial coordinate $r$. Within this ansatz, we calculate
the non-perturbative and non-singular black hole solutions that coincide
with the two-dimensional string black hole model described by Horne
and Horowitz in the perturbative limit $\alpha^{\prime}\rightarrow0$
\cite{Mandal:1991tz,Horne:1991gn}. This result indicates that the
properties of the event horizon remain unaffected by the $\alpha^{\prime}$
corrections in string theory, with only the curvature singularity
at $r=0$ being removed. Secondly, we adopt an ansatz derived from
Witten's two-dimensional black hole and it often used in the WZW model
\cite{Witten:1991}. Within this framework, we obtain two distinct
metrics that describe the regions inside and outside the event horizon.
In our previous letter \cite{Ying:2022xaj}, we successfully removed
the curvature singularity within the metric inside the event horizon
of the black hole. In this paper, we focus on studying the regular
solution outside the event horizon. The obtained result reveals that
the $\alpha^{\prime}$ corrections do not introduce any additional
curvature singularity outside the event horizon, and the properties
of the event horizon remain unaffected within this gauge. Moreover,
we demonstrate that the regular solutions between these two gauges
can be transformed into each other through appropriate coordinate
transformations.

The reminder of this paper is outlined as follows. In section 2, we
briefly review the two-dimensional string black hole in tree-level
bosonic string theory. In section 3, we calculate the non-perturbative
and non-singular black hole solutions in the Schwarzschild gauge.
In section 4, we focus on the the regular solutions in unitary gauge,
encompassing the metrics inside and outside the event horizon of the
black hole. Section 5 is a conclusion.

\section{Brief review of two-dimensional string black hole}

In this paper, our aim is to obtain regular solutions that precisely
solve the EOM of the complete $\alpha^{\prime}$ corrected closed
string theory. In the perturbative regime where $\alpha^{\prime}\rightarrow0$,
the solutions of two-dimensional string effective action reduces to
the traditional two-dimensional string black hole. Let us begin by
recalling the two-dimensional low energy effective action of the closed
string:

\begin{equation}
S=\int d^{2}x\sqrt{-g}e^{-2\phi}\left(R+4\left(\nabla\phi\right)^{2}+\lambda^{2}\right),\label{eq:original action}
\end{equation}

\noindent where $g_{\mu\nu}$ represents the string metric, $\phi$
is the physical dilaton, $\lambda^{2}=-\frac{2\left(D-26\right)}{3\alpha^{\prime}}$
is a constant and we set the Kalb-Ramond field $b_{\mu\nu}$ to zero
for simplicity. This two-dimensional gravitational theory exhibits
dynamics due to the pre-factor $e^{-2\phi}$. The black hole solution
derived form this action is given by \cite{Mandal:1991tz,Horne:1991gn}:

\begin{eqnarray}
ds^{2} & = & -\left(1-\frac{M}{r}\right)dt^{2}+\left(1-\frac{M}{r}\right)^{-1}\frac{1}{\lambda^{2}r^{2}}dr^{2},\nonumber \\
\phi & = & -\frac{1}{2}\ln\left(\frac{2}{M}r\right).\label{eq:3D black string metric}
\end{eqnarray}

\noindent where we set the integral constant $\phi_{0}$ of $\phi$
to zero for simplicity. In this metric (\ref{eq:3D black string metric}),
the event horizon is located at $r=M$, and the curvature singularity
occurs at $r=0$ due to the scalar curvature $R=\frac{\lambda^{2}M}{r}$.
The coordinate system of this metric is also known as ``Schwarzschild''
gauge. It is worth to note that there exist two types of coordinate
transformations, namely unitary gauge, cover different regions of
the maximally extended spacetime. The first one is given by:

\begin{equation}
\frac{r}{M}=\cosh^{2}\left(\frac{\lambda}{2}x\right),\label{eq:coordinates transformation}
\end{equation}

\noindent where $r\geq M$ and $x\geq0$. By utilizing this coordinate
transformation, the metric (\ref{eq:3D black string metric}) becomes

\begin{eqnarray}
ds^{2} & = & -\tanh^{2}\left(\frac{\lambda}{2}x\right)dt^{2}+dx^{2},\nonumber \\
\Phi & = & -\ln\left(\sinh\left(\lambda x\right)\right),\label{eq:O(2,2) metric}
\end{eqnarray}

\noindent where the $O\left(d,d\right)$ invariant dilaton is defined
as

\begin{equation}
\Phi=2\phi-\ln\sqrt{\det g_{ij}}.\label{eq:odd invariant dilaton}
\end{equation}

\noindent This metric is widely known as Witten's two-dimensional
black hole solution, which was obtained through the $SL\left(2,R\right)/U\left(1\right)$
gauged WZW model \cite{Witten:1991}. This metric can be directly
applied in the Hohm-Zwiebach action. However, it does not possess
a curvature singularity since it only describes the region outside
the event horizon ($x=0$). In this region, the scalar curvature $R_{0}=\lambda^{2}\cosh^{-2}\left(\frac{\lambda x}{2}\right)$
remains regular. To investigate the curvature singularity of the metric
(\ref{eq:3D black string metric}), we can employ the second type
of coordinate transformation:

\begin{equation}
\frac{r}{M}=\cos^{2}\left(\frac{\lambda}{2}x\right),\label{eq:coor trans 2}
\end{equation}

\noindent where $0\leq r\leq M$ and we only consider a single period,
namely $0\leq x\leq\frac{\pi}{\lambda}$. Using this transformation,
the metric (\ref{eq:3D black string metric}) takes the following
form:

\begin{eqnarray}
ds^{2} & = & -dx^{2}+\tan^{2}\left(\frac{\lambda}{2}x\right)dt^{2},\nonumber \\
\Phi & = & -\ln\left(\sin\left(\lambda x\right)\right).\label{eq:O(2,2) inner metric}
\end{eqnarray}

\noindent This metric describes the inner region of the black hole,
where $x$ plays the role of a time-like direction. The metric (\ref{eq:O(2,2) inner metric})
topologically corresponds to a disk, the event horizon located at
$x=0$ and the curvature singularity a single period the boundary
of the disk $x=\frac{\pi}{\lambda}$. The scalar curvature is given
by $R_{0}=\lambda^{2}\cos^{-2}\left(\frac{\lambda x}{2}\right)$.
Therefore, our aim is to remove the curvature singularities of the
(\ref{eq:3D black string metric}) and (\ref{eq:O(2,2) inner metric})
by incorporating the complete $\alpha^{\prime}$ corrections.

\section{Two-dimensional regular string black hole in Schwarzschild gauge}

In this section, we plan to remove the curvature singularity of two-dimensional
black hole in Schwarzschild gauge (\ref{eq:3D black string metric}).
In order to remove the singularity of the black hole solution (\ref{eq:3D black string metric})
by using Hohm-Zwiebach action, we introduce the following notations:

\begin{eqnarray}
ds^{2} & = & -n\left(r\right)^{2}dr^{2}+a\left(r\right)^{2}dt^{2},\nonumber \\
a\left(r\right)^{2} & = & -\left(1-\frac{M}{r}\right),\nonumber \\
n\left(r\right)^{2} & = & -\left(1-\frac{M}{r}\right)^{-1}\frac{1}{\lambda^{2}r^{2}},\nonumber \\
\Phi & = & -\log\left(2\sqrt{\frac{r}{M}\left(1-\frac{r}{M}\right)}\right),\label{eq:inner metric}
\end{eqnarray}

\noindent where $r$ effectively plays the role of a timelike direction,
and $n\left(r\right)$ can be seen as a lapse function. It should
be noted that the determinant $\sqrt{\det g_{ij}}$ does not include
$g_{rr}$ of the (\ref{eq:inner metric}). Due to this definition,
this metric describes the region $0\leq r\leq M$ for this definition.
Based on the solution (\ref{eq:inner metric}), we can extract the
ansatz:

\begin{equation}
ds^{2}=-n\left(r\right)^{2}dr^{2}+a\left(r\right)^{2}dt^{2}.\label{eq:ansatz}
\end{equation}

\noindent The closed string configurations that depend on this metric
possess an $O\left(1,1\right)$ symmetry. Using this ansatz, Hohm
and Zwiebach demonstrated that the following low energy effective
action with complete $\alpha^{\prime}$ corrections can be rewritten
as

\begin{eqnarray}
I_{HZ} & = & \int d^{2}x\sqrt{-g}e^{-2\phi}\left(R+4\left(\partial\phi\right)^{2}+\frac{1}{4}\alpha^{\prime}\left(R^{\mu\nu\rho\sigma}R_{\mu\nu\rho\sigma}+\ldots\right)+\alpha^{\prime2}\left(\ldots\right)+\ldots\right)\nonumber \\
 & = & \int dxe^{-\Phi}\left(-\frac{1}{n}\dot{\Phi}^{2}+\sum_{k=1}^{\infty}\left(-\alpha^{\prime}\right)^{k-1}\frac{2^{2k+1}}{n^{2k-1}}c_{k}H^{2k}\right),\label{eq:corrected action}
\end{eqnarray}

\noindent where the dot denotes $\dot{f}\left(r\right)\equiv\partial_{r}f\left(r\right)$,
$H\left(r\right)\equiv\frac{\dot{a}\left(r\right)}{a\left(r\right)}$,
$c_{1}=-\frac{1}{8}$, $c_{2}=\frac{1}{64}$, $c_{3}=-\frac{1}{3.2^{7}}$,
$c_{4}=\frac{1}{2^{15}}-\frac{1}{2^{12}}\zeta\left(3\right)$ and
$c_{k>4}$'s are unknown coefficients for the bosonic case \cite{Codina:2021cxh}.
In addition, only $a\left(r\right)$ and $\Phi$ contribute to the
$O(d,d)$ transformation:

\begin{equation}
\Phi\rightarrow\Phi,\qquad a\rightarrow a^{-1},\qquad H\rightarrow-H.
\end{equation}

\noindent Then, we can add an $O\left(d,d\right)$ invariant constant
into the action (\ref{eq:corrected action}) in order to include the
tree-level action (\ref{eq:original action}):

\begin{equation}
I_{m}=\int d^{2}xe^{-\Phi}\lambda^{2}.
\end{equation}

\noindent Therefore, EOM can be given by

\begin{eqnarray}
\frac{1}{n^{2}}\ddot{\Phi}-\frac{\dot{n}}{n^{3}}\dot{\Phi}+\frac{1}{2}\frac{1}{n}Hf\left(H\right) & = & 0,\nonumber \\
\frac{d}{dr}\left(e^{-\Phi}f\left(H\right)\right) & = & 0,\nonumber \\
\frac{1}{n^{2}}\dot{\Phi}^{2}+g\left(H\right)+\lambda^{2} & = & 0,\label{eq:corrected EOM}
\end{eqnarray}

\noindent with an extra constraint $n\dot{g}\left(H\right)=H\dot{f}\left(H\right)$
and

\begin{eqnarray}
f\left(H\right) & = & \sum_{\text{k=1}}^{\infty}\left(-\alpha^{\prime}\right)^{k-1}2^{2\left(k+1\right)}kc_{k}\left(\frac{1}{n}H\right)^{2k-1}=-2\frac{1}{n}H-\alpha^{\prime}2\frac{1}{n^{3}}H^{3}+\cdots,\nonumber \\
g\left(H\right) & = & \sum_{\text{k=1}}^{\infty}\left(-\alpha^{\prime}\right)^{k-1}2^{2k+1}\left(2k-1\right)c_{k}\left(\frac{1}{n}H\right)^{2k}=-\frac{1}{n^{2}}H^{2}-\alpha^{\prime}\frac{3}{2}\frac{1}{n^{4}}H^{4}+\cdots.\label{eq:EOM fh gh}
\end{eqnarray}

\noindent The solution (\ref{eq:3D black string metric}), also referred
to as (\ref{eq:inner metric}), is found to satisfy the EOM (\ref{eq:corrected EOM})
at the zeroth order of $\alpha^{\prime}$. It is important to emphasize
that $\alpha^{\prime}$ is an arbitrary positive constant within the
EOM (\ref{eq:corrected EOM}). Therefore, equations (\ref{eq:EOM fh gh})
do not represent a simple expansion as $\alpha^{\prime}\rightarrow0$;
rather, they form a non-perturbative series in $\alpha^{\prime}$.

To address the issue of the curvature singularity present in (\ref{eq:3D black string metric})
and achieve a non-perturbative and non-singular solution to the EOM
(\ref{eq:corrected EOM}) incorporating complete $\alpha^{\prime}$
corrections, we outline the following strategy:
\begin{enumerate}
\item Calculate the perturbative solutions to the EOM (\ref{eq:corrected EOM})
in an order-by-order manner with respect to $\alpha^{\prime}$.
\item Propose a non-perturbative and non-singular dilaton $\Phi$, which
covers the perturbative dilaton solution as $\alpha^{\prime}\rightarrow0$
in the initial step.
\item Since, the dilaton $\Phi$ governs all field solutions, we can obtain
the regular functions $f\left(H\right)$ and $g\left(H\right)$ from
the regular $\Phi$ using the EOM (\ref{eq:corrected EOM}).
\item Finally, we can obtain the regular Hubble parameter $H$ by solving
the constraint equation $n\dot{g}\left(H\right)=H\dot{f}\left(H\right)$.
\end{enumerate}
By following these steps, we obtain solutions that are regular and
fulfill the requirements of the EOM (\ref{eq:corrected EOM}).

We can now proceed to calculate the perturbative solutions of the
EOM (\ref{eq:corrected EOM}). For convenience, let us introduce a
new variable $\Omega$ defined as
\begin{equation}
\Omega\equiv e^{-\Phi},\label{eq:pertur notation}
\end{equation}

\noindent where $\dot{\Omega}=-\dot{\Phi}\Omega$ and $\ddot{\Omega}=\left(-\ddot{\Phi}+\dot{\Phi}^{2}\right)\Omega$.
This allows us to rewrite the EOM (\ref{eq:corrected EOM}) in terms
of $\Omega$ as follows:

\noindent
\begin{eqnarray}
\ddot{\Omega}-\frac{\dot{n}}{n}\dot{\Omega}-\left(h\left(H\right)-\lambda^{2}\right)\Omega n^{2} & = & 0,\nonumber \\
\frac{d}{dr}\left(\Omega f\left(H\right)\right) & = & 0,\nonumber \\
\dot{\Omega}^{2}+\left(g\left(H\right)+\lambda^{2}\right)\Omega^{2}n^{2} & = & 0,\label{eq:reEOM}
\end{eqnarray}

\noindent where we define a new function

\begin{equation}
h\left(H\right)\equiv\frac{1}{2}\frac{1}{n}Hf\left(H\right)-g\left(H\right)=\alpha^{\prime}\frac{1}{2}\frac{1}{n^{4}}H^{4}+\ldots,
\end{equation}

\noindent It is easy to see that $h\left(H\right)=0$ at the zeroth
order of $\alpha^{\prime}$. Moving forward, we assume that the perturbative
solutions of the EOM (\ref{eq:reEOM}):

\begin{eqnarray}
\Omega\left(r\right) & = & \Omega_{0}\left(r\right)+\alpha^{\prime}\Omega_{1}\left(r\right)+\alpha^{\prime2}\Omega_{2}\left(r\right)+\ldots,\nonumber \\
H\left(r\right) & = & H_{0}\left(r\right)+\alpha^{\prime}H_{1}\left(r\right)+\alpha^{\prime2}H_{2}\left(r\right)+\ldots.\nonumber \\
n\left(r\right) & = & n_{0}\left(r\right),\label{eq:pertur form}
\end{eqnarray}

\noindent where $\Omega_{i}$ and $H_{i}$ represent the $i$-th order
perturbative solutions. Moreover, we have two reasons for assuming
that $n\left(r\right)$ does not receive $\alpha^{\prime}$ corrections
in (\ref{eq:pertur form}): i) It is straightforward to determine
the exact solutions of the EOM (\ref{eq:reEOM}); ii) We aim to ensure
that the properties of the event horizon ($g_{rr}\left(r_{H}\right)=n\left(r_{H}\right)=0$)
remain unaltered by $\alpha^{\prime}$ corrections. By substituting
the perturbative forms (\ref{eq:pertur form}), the functions $h\left(H\right)$,
$f\left(H\right)$ and $g\left(H\right)$ become

\begin{eqnarray}
h\left(H\right) & = & \alpha^{\prime}\frac{1}{2}\frac{1}{n_{0}^{4}}H_{0}^{4}+\ldots,\nonumber \\
f\left(H\right) & = & -2\frac{1}{n_{0}}H_{0}-\alpha^{\prime}\left(2\frac{1}{n_{0}^{3}}H_{0}^{3}+2\frac{1}{n_{0}}H_{1}\right)+\cdots,\nonumber \\
g\left(H\right) & = & -\frac{1}{n_{0}^{2}}H_{0}^{2}-\alpha^{\prime}\left(\frac{3}{2}\frac{1}{n_{0}^{4}}H_{0}^{4}+2\frac{1}{n_{0}^{2}}H_{0}H_{1}\right)+\cdots.\nonumber \\
\end{eqnarray}

\noindent We can then substitute these expansions back into the EOM
(\ref{eq:reEOM}) and solve the resulting differential equations at
each order of $\alpha^{\prime}$ to obtain the expressions for $\Omega_{i}$
and $H_{i}$. For instance, the EOM (\ref{eq:reEOM}) at the zeroth
order of $\alpha^{\prime}$ yield:

\begin{eqnarray}
\ddot{\Omega}_{0}-\frac{\dot{n}_{0}}{n_{0}}\dot{\Omega}_{0}+\lambda^{2}\Omega_{0}n_{0}^{2} & = & 0,\nonumber \\
\frac{d}{dr}\left(-2\Omega_{0}\frac{1}{n_{0}}H_{0}\right) & = & 0,\nonumber \\
\dot{\Omega}_{0}^{2}-\left(\frac{1}{n_{0}^{2}}H_{0}^{2}-\lambda^{2}\right)\Omega_{0}^{2}n_{0}^{2} & = & 0,\label{eq:0th EOM}
\end{eqnarray}

\noindent As anticipated, we obtained the following solution:

\begin{eqnarray}
\Omega_{0}\left(r\right) & = & 2\frac{\sqrt{r\left(M-r\right)}}{M},\nonumber \\
H_{0}\left(r\right) & = & -\frac{M}{2Mr-2r^{2}},\nonumber \\
n_{0}\left(r\right) & = & \left(\frac{M}{r}-1\right)^{-1/2}\frac{1}{\lambda r},
\end{eqnarray}

\noindent which covers the solutions (\ref{eq:3D black string metric})
or (\ref{eq:inner metric}). Subsequently, the EOM (\ref{eq:reEOM})
at the first order of $\alpha^{\prime}$ yield:

\begin{eqnarray}
\ddot{\Omega}_{1}-\frac{\dot{n}_{0}}{n_{0}}\dot{\Omega}_{1}+\lambda^{2}n_{0}^{2}\Omega_{1}-\frac{1}{2}\frac{H_{0}^{4}}{n_{0}^{2}}\Omega_{0} & = & 0,\nonumber \\
2\frac{H_{0}}{n_{0}}\Omega_{1}+2\left(\frac{H_{0}(r){}^{3}}{n_{0}(r){}^{3}}+\frac{H_{1}}{n_{0}}\right)\Omega_{0} & = & 0,\nonumber \\
2\dot{\Omega}_{0}\dot{\Omega}_{1}+2\left(\lambda^{2}n_{0}^{2}-H_{0}^{2}\right)\Omega_{0}\Omega_{1}-\left(\frac{3H_{0}^{4}}{2n_{0}^{2}}+2H_{0}H_{1}\right)\Omega_{0}^{2} & = & 0.
\end{eqnarray}

\noindent The solution is given by:

\begin{eqnarray}
\Omega_{1}\left(r\right) & = & \frac{\lambda^{2}\left(M^{2}-8Mr+8r^{2}\right)}{8M\sqrt{r\left(M-r\right)}},\nonumber \\
H_{1}\left(r\right) & = & \frac{M\left(5M^{2}-8Mr+8r^{2}\right)\lambda^{2}}{32r^{2}\left(M-r\right)^{2}}.
\end{eqnarray}

\noindent Therefore, the perturbative solution, including the first
two orders of $\alpha^{\prime}$ is given by:

\begin{eqnarray}
H\left(r\right) & = & -\frac{M}{2Mr-2r^{2}}+\frac{M\left(5\lambda^{2}M^{2}-8\lambda^{2}Mr+8\lambda^{2}r^{2}\right)}{32r^{2}\left(M-r\right)^{2}}\alpha^{\prime}+\cdots\nonumber \\
\Omega\left(r\right) & = & 2\frac{\sqrt{r\left(M-r\right)}}{M}+\frac{\lambda^{2}\left(M^{2}-8Mr+8r^{2}\right)}{8M\sqrt{r\left(M-r\right)}}\alpha^{\prime}+\cdots,\label{eq:perturbed solution}
\end{eqnarray}

\noindent From equation (\ref{eq:pertur notation}), we have

\begin{eqnarray}
\Phi\left(r\right) & = & -\frac{1}{2}\log\left(\frac{4r\left(M-r\right)}{M^{2}}\right)-\frac{\lambda^{2}\left(M^{2}-8Mr+8r^{2}\right)}{16r\left(M-r\right)}\alpha^{\prime}+\cdots.\label{eq:perturbed solution dilaton}
\end{eqnarray}

\begin{figure}[H]
\begin{centering}
\includegraphics[scale=0.4]{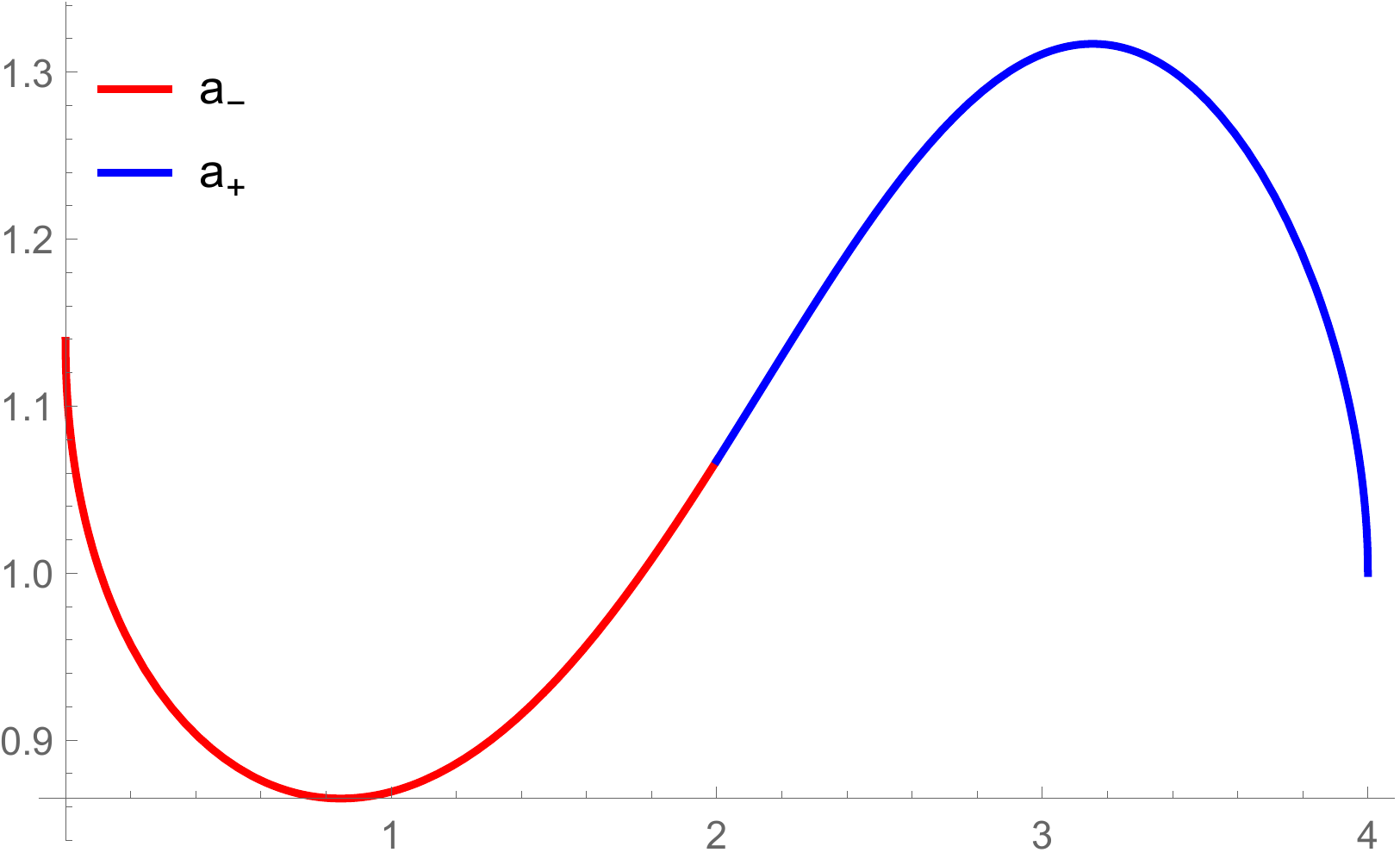}
\par\end{centering}
\centering{}\caption{\label{fig:a} The figure of $a_{\pm}\left(r\right)$.}
\end{figure}

\noindent Based on the perturbative solutions (\ref{eq:perturbed solution})
and (\ref{eq:perturbed solution dilaton}), with some trial and error,
we deduced a non-perturbative and non-singular solution:

\begin{eqnarray}
\Phi\left(r\right) & = & -\frac{1}{2}\log\left(\frac{\alpha^{\prime}\lambda^{2}}{2}-\frac{4r\left(\alpha^{\prime}\lambda^{2}-1\right)\left(M-r\right)}{M^{2}}\right),\qquad n\left(r\right)=\left(\frac{M}{r}-1\right)^{-1/2}\frac{1}{\lambda r},\nonumber \\
H\left(r\right) & = & -\frac{\sqrt{2}M\left(\alpha^{\prime}\lambda^{2}-1\right)\left(\alpha^{\prime}\lambda^{2}M^{2}-8Mr+8r^{2}\right)}{\sqrt{r\left(M-r\right)}\left(\alpha^{\prime}\lambda^{2}\left(M^{2}-8Mr+8r^{2}\right)+8r\left(M-r\right)\right)^{3/2}},\nonumber \\
f\left(r\right) & = & 2\lambda\left(\frac{\alpha^{\prime}\lambda^{2}}{2}-\frac{4r\left(\alpha^{\prime}\lambda^{2}-1\right)\left(M-r\right)}{M^{2}}\right)^{-1/2},\nonumber \\
g\left(r\right) & = & \frac{\lambda^{2}M^{2}\left(-\alpha^{\prime2}\lambda^{4}M^{2}+16\alpha^{\prime}\lambda^{2}r\left(M-r\right)+16r\left(r-M\right)\right)}{\left(\alpha^{\prime}\lambda^{2}\left(M^{2}-8Mr+8r^{2}\right)+8r\left(M-r\right)\right)^{2}},\label{eq:r solution}
\end{eqnarray}

\noindent which satisfies the EOM (\ref{eq:corrected EOM}) and exhibits
a perfect match with the perturbative solutions (\ref{eq:perturbed solution}),
(\ref{eq:perturbed solution dilaton}) and (\ref{eq:EOM fh gh}) in
the limit of $\alpha^{\prime}\rightarrow0$. From the derived solution
(\ref{eq:r solution}), the metric takes the form:

\begin{equation}
ds^{2}=a_{\pm}\left(r\right)^{2}dt^{2}+\left(1-\frac{M}{r}\right)^{-1}\frac{1}{\lambda^{2}r^{2}}dr^{2},
\end{equation}

\noindent where the additional minus sign in $a\left(r\right)^{2}$
originates from our ansatz. The functions $a_{\pm}\left(r\right)$
are given by

\begin{eqnarray}
a_{\pm}\left(r\right) & = & \exp\sqrt{2}\left[\sqrt{\frac{1}{\alpha^{\prime}\lambda^{2}}}\mathbb{F}\left(\cos^{-1}\left(\frac{2r}{M}-1\right)|2-\frac{2}{\alpha^{\prime}\lambda^{2}}\right)-\sqrt{\alpha^{\prime}\lambda^{2}}\mathbb{E}\left(\cos^{-1}\left(\frac{2r}{M}-1\right)|2-\frac{2}{\alpha^{\prime}\lambda^{2}}\right)\right.\nonumber \\
 &  & \left.\pm\frac{4\left(\alpha^{\prime}\lambda^{2}-1\right)\sqrt{r\left(M-2r\right)^{2}\left(M-r\right)}}{M\sqrt{\alpha^{\prime}\lambda^{2}\left(M^{2}-8Mr+8r^{2}\right)+8r\left(M-r\right)}}\right],
\end{eqnarray}

\noindent where $\mathbb{F}\left(\phi|m\right)$ and $\mathbb{E}\left(\phi|m\right)$
represent elliptic integrals of the first and second kinds. Furthermore,
$a_{+}\left(r\right)$ is valid in the region $\frac{M}{2}\leq r\leq M$,
while $a_{-}\left(r\right)$ is valid in the region $0\leq r\leq\frac{M}{2}$,
with continuity at $\frac{M}{2}$. It is easy to see that the event
horizon remains unaffected by the $\alpha^{\prime}$ corrections.
Figure (\ref{fig:a}) displays the plot $a_{\pm}\left(r\right)$ using
the values $\alpha^{\prime}\lambda^{2}=4/3$ and $M=4$.

\noindent To examine the presence of a physical singularity, we evaluate
the Kretschmann scalar:

\begin{eqnarray}
 &  & R_{\mu\nu\rho\sigma}R^{\mu\nu\rho\sigma}=\frac{1}{2}R_{\mu\nu}R^{\mu\nu}=R^{2}\nonumber \\
 &  & =\lambda^{4}\left(2r\left(M-r\right)\left(\dot{H}+H^{2}\right)+\left(M-2r\right)H\right)^{2}\nonumber \\
 &  & =\frac{\lambda^{4}M^{2}\left(1-\alpha^{\prime}\lambda^{2}\right)^{2}}{A\left(r\right)^{6}}\left(-16C\left(r\right)A\left(r\right)^{3/2}-4M\left(\alpha^{\prime}\lambda^{2}-1\right)B\left(r\right)^{2}+24\left(\alpha^{\prime}\lambda^{2}-1\right)C\left(r\right)B\left(r\right)A\left(r\right)^{1/2}\right)^{2},\nonumber \\
\end{eqnarray}

\noindent where

\begin{eqnarray}
A\left(r\right) & = & \alpha^{\prime}\lambda^{2}\left(M^{2}-8Mr+8r^{2}\right)+8r\left(M-r\right),\nonumber \\
B\left(r\right) & = & \alpha^{\prime}\lambda^{2}M^{2}-8Mr+8r^{2},\nonumber \\
C\left(r\right) & = & \sqrt{2}\left(M-2r\right)\sqrt{r\left(M-r\right)}.
\end{eqnarray}

\begin{figure}[H]
\begin{centering}
\includegraphics[scale=0.4]{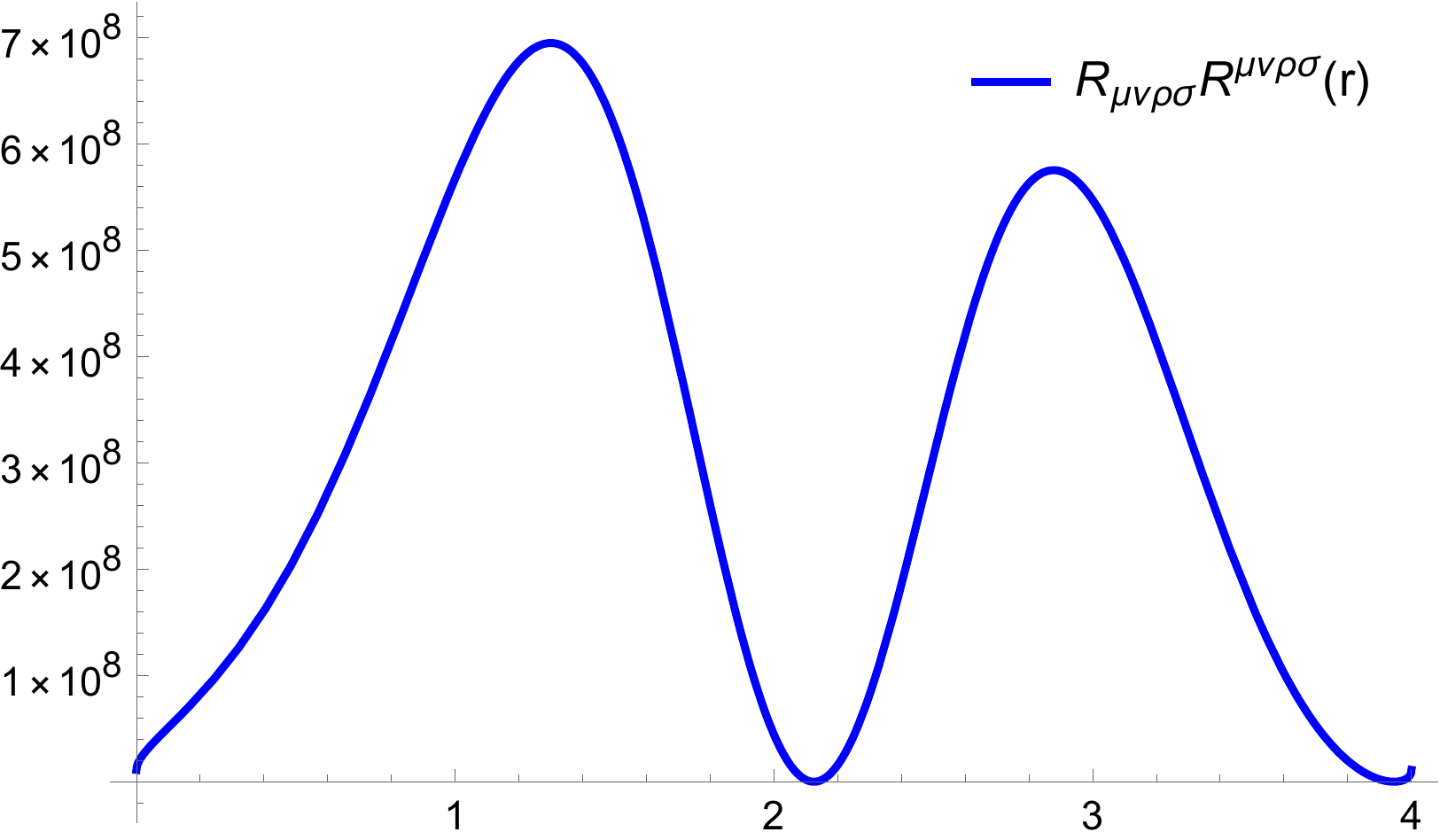}
\par\end{centering}
\centering{}\caption{\label{fig:scalar curvature} The figure of Kretschmann scalar $0\protect\leq r\protect\leq M$,
where we set $\alpha^{\prime}\lambda^{2}=\frac{4}{3}$, $\lambda=100$
and $M=4$.}
\end{figure}

\noindent Therefore, the solutions (\ref{eq:r solution}) are regular
only when $A\left(r\right)=0$ has no any real root. This condition
leads to the following result:

\begin{equation}
1\leq\alpha^{\prime}\lambda^{2}<2,
\end{equation}

\noindent which is consistent with the result $\lambda^{2}\sim\frac{1}{\alpha^{\prime}}$
(cosmological constant problem in string theory) within the framework
of string theory. The Kretschmann scalar is displayed in Figure (\ref{fig:scalar curvature}).

In summary, the key results of this subsection are as follows:
\begin{itemize}
\item The $\alpha^{\prime}$ corrections eliminate the curvature singularity
located at $r=0$.
\item The $\alpha^{\prime}$ corrections do not affect the event horizon,
as it is maintained by $n\left(r\right)=\left(\frac{M}{r}-1\right)^{-1/2}\frac{1}{\lambda r}$
.
\end{itemize}

\section{Two-dimensional regular string black hole in unitary gauge}

In this section, we examine the regular black hole solution within
the ansatz (\ref{eq:O(2,2) metric}) and (\ref{eq:O(2,2) inner metric}).
These two ansatz cover the inside and outside regions of the event
horizon of two-dimensional black hole separately.

\subsection{Inside the event horizon}

The solutions inside the event horizon were previously derived in
ref. \cite{Ying:2022xaj}, and we provide the main results here. Recalling
the two-dimensional black hole metric inside the event horizon:

\begin{eqnarray}
ds^{2} & = & -dx^{2}+\tan^{2}\left(\frac{\lambda}{2}x\right)dt^{2},\nonumber \\
\Phi & = & -\ln\left(\sin\left(\lambda x\right)\right),
\end{eqnarray}

\noindent we can extract the ansatz:

\begin{equation}
ds^{2}=-dx^{2}+a_{in}\left(x\right)^{2}dt^{2}.
\end{equation}

\noindent In this background, the EOM of Hohm-Zwiebach action are
given by:

\begin{eqnarray}
\ddot{\Phi}_{in}+\frac{1}{2}H_{in}f\left(H_{in}\right) & = & 0,\nonumber \\
\frac{d}{dx}\left(e^{-\Phi_{in}}f\left(H_{in}\right)\right) & = & 0,\nonumber \\
\dot{\Phi}_{in}^{2}+g\left(H_{in}\right)+\lambda^{2} & = & 0,\label{eq:corrected EOM-2}
\end{eqnarray}

\noindent where

\begin{eqnarray}
f\left(H_{in}\right) & = & \sum_{k=1}^{\infty}\left(-\alpha^{\prime}\right)^{k-1}2^{2\left(k+1\right)}kc_{k}H_{in}^{2k-1}=-2H_{in}-\alpha^{\prime}2H_{in}^{3}+\cdots,\nonumber \\
g\left(H_{in}\right) & = & \sum_{k=1}^{\infty}\left(-\alpha^{\prime}\right)^{k-1}2^{2k+1}\left(2k-1\right)c_{k}H_{in}^{2k}=-H_{in}^{2}-\alpha^{\prime}\frac{3}{2}H_{in}^{4}+\cdots,\label{eq:EOM fh gh-2}
\end{eqnarray}

\noindent and there is an additional constraint: $\dot{g}\left(H_{in}\right)=H_{in}\dot{f}\left(H_{in}\right)$.
Since the ansatz is analogous to the cosmological background, we denote
$\dot{f}\left(x\right)\equiv\partial_{x}f\left(x\right)$, $H_{in}\left(x\right)\equiv\frac{\dot{a}_{in}\left(x\right)}{a_{in}\left(x\right)}$,
$c_{1}=-\frac{1}{8}$, $c_{2}=\frac{1}{64}$, $c_{3}=-\frac{1}{3.2^{7}}$,
$c_{4}=\frac{1}{2^{15}}-\frac{1}{2^{12}}\zeta\left(3\right)$ and
$c_{k>4}$'s are unknown coefficients for the bosonic case \cite{Codina:2021cxh}.
The regular solutions are given by:

\begin{eqnarray}
\Phi_{in}\left(x\right) & = & \log\sqrt{\frac{1+\alpha^{\prime}\lambda^{2}}{\sin^{2}\left(\lambda x\right)+\frac{1}{2}\alpha^{\prime}\lambda^{2}\sin^{2}\left(\lambda x\right)\left(\cot^{2}\left(\lambda x\right)+1\right)}},\nonumber \\
H_{in}\left(x\right) & = & -\frac{\sqrt{2}\lambda\left(\left(\alpha^{\prime}\lambda^{2}+1\right)\cos\left(2\lambda x\right)-1\right)}{\left(\alpha^{\prime}\lambda^{2}+1\right)^{1/2}\left(\alpha^{\prime}\lambda^{2}+1-\cos\left(2\lambda x\right)\right)^{3/2}},\nonumber \\
f_{in}\left(x\right) & = & -2\sqrt{2}\lambda\left(\frac{\alpha^{\prime}\lambda^{2}+1}{\alpha^{\prime}\lambda^{2}+1-\cos\left(2\lambda x\right)}\right)^{1/2},\nonumber \\
g_{in}\left(x\right) & = & \frac{\lambda^{2}}{\left(\alpha^{\prime}\lambda^{2}+1-\cos\left(2\lambda x\right)\right)^{2}}\left(-\alpha^{\prime}\lambda^{2}\left(\alpha^{\prime}\lambda^{2}+2\right)+2\left(\alpha^{\prime}\lambda^{2}+1\right)\cos\left(2\lambda x\right)-2\right),\label{eq:solution}
\end{eqnarray}

\noindent and the corresponding scale factor is given by:

\begin{equation}
ds^{2}=a_{in}\left(x\right)^{2}dt^{2}-dx^{2}.\label{eq:inside solution}
\end{equation}

\noindent with

\begin{eqnarray}
a_{in}\left(x\right) & = & C_{in}\exp\sqrt{2}\left[\sqrt{\frac{\alpha^{\prime}\lambda^{2}+1}{\alpha^{\prime}\lambda^{2}}}\mathbb{F}\left(x\lambda\left|-\frac{2}{\alpha^{\prime}\lambda^{2}}\right.\right)\right.\nonumber \\
 &  & \left.-\sqrt{\frac{\alpha^{\prime}\lambda^{2}}{\alpha^{\prime}\lambda^{2}+1}}\mathbb{E}\left(x\lambda\left|-\frac{2}{\alpha^{\prime}\lambda^{2}}\right.\right)-\frac{\sin\left(2\lambda x\right)}{\sqrt{\left(\alpha^{\prime}\lambda^{2}+1\right)\left(\alpha^{\prime}\lambda^{2}+1-\cos\left(2\lambda x\right)\right)}}\right],\label{eq:a solution}
\end{eqnarray}

\noindent where $\mathbb{F}\left(\phi|m\right)$ and $\mathbb{E}\left(\phi|m\right)$
are elliptic integrals of the first and second kinds, and $C_{in}$
is an integral constant. This regular solution has been studied in
ref. \cite{Ying:2022xaj}.

Finally, we wish to verify that by using the appropriate coordinate
transformation (\ref{eq:coor trans 2}) on (\ref{eq:solution}), we
can obtain the expected solution (\ref{eq:r solution})

\begin{eqnarray}
H\left(r\right) & = & \frac{\frac{dx}{dr}\frac{d}{dx}a_{in}\left(x\right)}{a_{in}\left(x\right)}=\frac{dx}{dr}H\left(x\right)\nonumber \\
 & = & -\frac{1}{\lambda M\sqrt{\frac{r}{M}-\frac{r^{2}}{M^{2}}}}\frac{\sqrt{2}\lambda\left(\alpha^{\prime}\lambda^{2}-1\right)\left(\alpha^{\prime}\lambda^{2}-1+\cos\left(2\lambda x\right)\right)}{\left(\left(\alpha^{\prime}\lambda^{2}-1\right)\cos\left(2\lambda x\right)+1\right)^{3/2}}.
\end{eqnarray}

\noindent Then, using the coordinate transformation (\ref{eq:coor trans 2}):
$r/M=\cos^{2}\left(\lambda x/2\right)$, we have

\begin{equation}
H\left(r\right)=-\frac{\sqrt{2}M\left(\alpha\lambda^{2}-1\right)\left(\alpha\lambda^{2}M^{2}-8Mr+8r^{2}\right)}{\sqrt{r(M-r)}\left(\alpha\lambda^{2}\left(M^{2}-8Mr+8r^{2}\right)+8r(M-r)\right)^{3/2}},
\end{equation}

\noindent which agrees with the solution (\ref{eq:r solution}).

In summary, the following results are obtained in this subsection:
\begin{itemize}
\item The curvature singularity located at $x=\pi/\lambda$ can be removed
by the $\alpha^{\prime}$ corrections.
\item The solution in unitary gauge (\ref{eq:solution}) agrees with the
solution in Schwarzschild gauge (\ref{eq:r solution}) using the appropriate
coordinate transformation (\ref{eq:coor trans 2}).
\end{itemize}

\subsection{Outside the event horizon}

Firstly, let us consider the metric that describes the region outside
the event horizon:

\begin{eqnarray}
ds^{2} & = & -\tanh^{2}\left(\frac{\lambda}{2}x\right)dt^{2}+dx^{2},\nonumber \\
\Phi & = & -\ln\left(\sinh\left(\lambda x\right)\right).
\end{eqnarray}

\noindent To apply the Hohm-Zwiebach action, we extract the ansatz
from the above metric, which is given by:

\begin{equation}
ds^{2}=dx^{2}-a_{out}\left(x\right)^{2}dt^{2}.
\end{equation}

\noindent This ansatz can be seen as the double wick rotation of the
cosmological background, and we use ``bar'' to denote this situation
in the following calculations. The corresponding Hohm-Zwiebach action
can be expressed as:

\begin{eqnarray}
\bar{I}_{HZ} & = & \int d^{3}x\sqrt{-g}e^{-2\phi}\left(R+4\left(\partial\phi\right)^{2}+\frac{1}{4}\alpha^{\prime}\left(R^{\mu\nu\rho\sigma}R_{\mu\nu\rho\sigma}+\ldots\right)+\alpha^{\prime2}\left(\ldots\right)+\ldots\right)\nonumber \\
 & = & -\int dxe^{-\Phi}\left(-\Phi^{\prime2}-\sum_{\text{k=1}}^{\infty}\left(-\alpha^{\prime}\right)^{k-1}2^{2k+1}\bar{c}_{k}H_{out}^{2k}\right),\label{eq:Odd action with alpha}
\end{eqnarray}

\noindent where $H_{out}\left(x\right)=\frac{\partial_{x}a_{out}\left(x\right)}{a_{out}\left(x\right)}$
and $A^{\prime}\left(x\right)\equiv\partial_{x}A\left(x\right)$.
It is important to note the difference in sign between the actions
in (\ref{eq:corrected action}) and (\ref{eq:Odd action with alpha})
for the two different ansatz. Moreover, we have $\bar{c}_{1}=c_{1}=-\frac{1}{8}$,
$\bar{c}_{2}=-c_{2}=-\frac{1}{64}$, $\bar{c}_{2k-1}=c_{2k-1}$, $\bar{c}_{2k}=-c_{2k}$.
Including the cosmological constant term:

\begin{equation}
\bar{I}_{m}=\int d^{3}xe^{-\Phi}\lambda^{2},
\end{equation}

\noindent the corresponding EOM for the action can be expressed as:

\begin{eqnarray}
\Phi^{\prime\prime}+\frac{1}{2}H_{out}f\left(H_{out}\right) & = & 0,\nonumber \\
\frac{d}{dx}\left(e^{-\Phi_{out}}f\left(H_{out}\right)\right) & = & 0,\nonumber \\
\Phi_{out}^{\prime2}+g\left(H_{out}\right)-\lambda^{2} & = & 0,\label{eq:corrected EOM-1}
\end{eqnarray}

\noindent where there is a constraint $g^{\prime}\left(H_{out}\right)=H_{out}f^{\prime}\left(H_{out}\right)$
and

\begin{eqnarray}
f\left(H_{out}\right) & = & \sum_{\text{k=1}}^{\infty}\left(-\alpha^{\prime}\right)^{k-1}2^{2\left(k+1\right)}k\bar{c}_{k}H_{out}^{2k-1}=-2H_{out}+\alpha^{\prime}2H_{out}^{3}+\cdots,\nonumber \\
g\left(H_{out}\right) & = & \sum_{\text{k=1}}^{\infty}\left(-\alpha^{\prime}\right)^{k-1}2^{2k+1}\left(2k-1\right)\bar{c}_{k}H_{out}^{2k}=-H_{out}^{2}+\alpha^{\prime}\frac{3}{2}H_{out}^{4}+\cdots.\label{eq:EOM fh gh-1}
\end{eqnarray}

\noindent To calculate the perturbative solutions, we also introduce
the variable $\Omega_{out}\equiv e^{-\Phi_{out}}$, $\Omega_{out}^{\prime}=-\Phi_{out}^{\prime}\Omega_{out}$
and $\Omega_{out}^{\prime\prime}=\left(-\Phi_{out}^{\prime\prime}+\Phi_{out}^{\prime2}\right)\Omega_{out}$.
The EOM (\ref{eq:corrected EOM-1}) can be written as follows:

\noindent
\begin{eqnarray}
\Omega_{out}^{\prime\prime}-\left(h\left(H_{out}\right)+\lambda^{2}\right)\Omega_{out} & = & 0,\nonumber \\
\frac{d}{dx}\left(\Omega_{out}f\left(H_{out}\right)\right) & = & 0,\nonumber \\
\Omega_{out}^{\prime2}+\left(g\left(H_{out}\right)-\lambda^{2}\right)\Omega_{out}^{2} & = & 0,\label{eq:reEOM-1}
\end{eqnarray}

\noindent where we define a new function

\begin{equation}
h\left(H_{out}\right)\equiv\frac{1}{2}H_{out}f\left(H_{out}\right)-g\left(H_{out}\right)=-\alpha^{\prime}\frac{1}{2}H_{out}^{4}+\ldots,
\end{equation}

\noindent It is easy to see that $\bar{h}\left(\bar{H}\right)=0$
at the zeroth order of $\alpha^{\prime}$. We assume the perturbative
solutions of the EOM (\ref{eq:reEOM-1}) take the following forms:

\begin{eqnarray}
\Omega_{out}\left(x\right) & = & \Omega_{0}\left(x\right)+\alpha^{\prime}\Omega_{1}\left(x\right)+\alpha^{\prime2}\Omega_{2}\left(x\right)+\ldots,\nonumber \\
H_{out}\left(x\right) & = & H_{0}\left(x\right)+\alpha^{\prime}H_{1}\left(x\right)+\alpha^{\prime2}H_{2}\left(x\right)+\ldots.
\end{eqnarray}

\noindent where $\Omega_{i}$ and $\bar{H}_{i}$ represent the $i$-th
order of the perturbative solutions. Using these perturbative forms,
the functions $h\left(H_{out}\right)$, $f\left(H_{out}\right)$ and
$g\left(H_{out}\right)$ can be expressed as:

\begin{eqnarray}
h\left(H_{out}\right) & = & -\alpha^{\prime}\frac{1}{2}H_{0}^{4}+\ldots,\nonumber \\
f\left(H_{out}\right) & = & -2H_{0}+\alpha^{\prime}\left(2H_{0}^{3}-2H_{1}\right)+\cdots,\nonumber \\
g\left(H_{out}\right) & = & -H_{0}^{2}+\alpha^{\prime}\left(\frac{3}{2}H_{0}^{4}-2H_{0}H_{1}\right)+\cdots.
\end{eqnarray}

\noindent By substituting these perturbative forms back into the EOM
(\ref{eq:reEOM-1})and solving the resulting differential equations
at each order of $\alpha^{\prime}$, we obtain $\Omega_{i}$ and $\bar{H}_{i}$.
At the zeroth order of $\alpha^{\prime}$ , the EOM (\ref{eq:reEOM-1})
gives the expected solution:

\begin{eqnarray}
\Omega_{0}^{\prime\prime}-\lambda^{2}\Omega_{0} & = & 0,\nonumber \\
\frac{d}{dx}\left(-2\Omega_{0}\bar{H}_{0}\right) & = & 0,\nonumber \\
\Omega_{0}^{\prime2}-\left(H_{0}^{2}+\lambda^{2}\right)\Omega_{0}^{2} & = & 0,\label{eq:0th EOM-1}
\end{eqnarray}

\noindent As expected, the zeroth order solution is given by:

\begin{eqnarray}
\Omega_{0} & = & \sinh\left(\lambda x\right),\nonumber \\
H_{0} & = & \lambda\text{csch}\left(\lambda x\right),
\end{eqnarray}

\noindent which corresponds to the solution (\ref{eq:O(2,2) metric}).
Moving on to the first order of $\alpha^{\prime}$, the EOM (\ref{eq:reEOM-1})
yield:

\begin{eqnarray}
\alpha^{\prime}\Omega_{1}^{\prime\prime}+\alpha^{\prime}\left(\frac{1}{2}H_{0}^{4}\Omega_{0}-\lambda^{2}\Omega_{1}\right) & = & 0,\nonumber \\
\alpha^{\prime}\left(-2H_{0}\Omega_{1}+\left(2H_{0}^{3}-2H_{1}\right)\Omega_{0}\right) & = & 0,\nonumber \\
2\alpha^{\prime}\Omega_{0}^{\prime}\Omega_{1}^{\prime}-2\alpha^{\prime}\left(H_{0}^{2}+\lambda^{2}\right)\Omega_{0}\Omega_{1}+\alpha^{\prime}\left(\frac{3}{2}H_{0}^{4}-2H_{0}H_{1}\right)\Omega_{0}^{2} & = & 0.
\end{eqnarray}

\noindent The solution is given by:

\begin{eqnarray}
\Omega_{1}\left(x\right) & = & -\frac{\lambda^{2}}{4}\mathrm{csch}\left(\lambda x\right),\nonumber \\
H_{1}\left(x\right) & = & \frac{\lambda^{3}}{4}\left(4+\cosh\left(2\lambda x\right)\right)\mathrm{csch}^{3}\left(\lambda x\right).
\end{eqnarray}

\noindent Therefore, the perturbative solution up to the first two
orders of $\alpha^{\prime}$ is given by:

\begin{eqnarray}
H_{out}\left(x\right) & = & \lambda\text{csch}\left(\lambda x\right)+\frac{\lambda^{3}}{4}\left(4+\cosh\left(2\lambda x\right)\right)\mathrm{csch}^{3}\left(\lambda x\right)\alpha^{\prime}+\cdots,\nonumber \\
\Omega_{out}\left(x\right) & = & \sinh\left(\lambda x\right)-\frac{\lambda^{2}}{4}\mathrm{csch}\left(\lambda x\right)\alpha^{\prime}+\cdots,\label{eq:perturbed solution-1}
\end{eqnarray}

\noindent Using the notation in (\ref{eq:pertur notation}), we can
express $\Phi_{out}\left(x\right)$ as

\begin{equation}
\Phi_{out}\left(x\right)=-\log\left(\sinh\left(\lambda x\right)\right)+\frac{\lambda^{2}}{4}\mathrm{csch}^{2}\left(\lambda x\right)\alpha^{\prime}+\cdots.\label{eq:perturbed solution dilaton-1}
\end{equation}

\noindent Based on the perturbative solutions (\ref{eq:perturbed solution-1})
and (\ref{eq:perturbed solution dilaton-1}), and after numerous trial
and error attempts, we can determine a non-perturbative and non-singular
solution:

\begin{eqnarray}
\Phi_{out}\left(x\right) & = & -\frac{1}{2}\log\left[\frac{2\sinh^{2}\left(\lambda x\right)}{\alpha^{\prime}\lambda^{2}\left(\coth^{2}\left(\lambda x\right)-1\right)+2}\right],\nonumber \\
H_{out}\left(x\right) & = & \frac{\sqrt{2}\lambda\left(\alpha^{\prime}\lambda^{2}\left(4\alpha^{\prime}\lambda^{2}-7\right)+\left(6\alpha^{\prime}\lambda^{2}-4\right)\cosh\left(2\lambda x\right)+\left(\alpha^{\prime}\lambda^{2}+1\right)\cosh\left(4\lambda x\right)+3\right)}{\left(\alpha^{\prime}\lambda^{2}+2\right)\left(\alpha^{\prime}\lambda^{2}+\cosh\left(2\lambda x\right)-1\right)^{5/2}},\nonumber \\
f_{out}\left(x\right) & = & -\frac{\lambda\left(\alpha^{\prime}\lambda^{2}+2\right)\text{csch}^{2}\left(\lambda x\right)\sqrt{\alpha^{\prime}\lambda^{2}+\cosh\left(2\lambda x\right)-1}}{\sqrt{2}},\nonumber \\
g_{out}\left(x\right) & = & -\frac{\lambda^{2}\left(3\alpha^{\prime}\lambda^{2}\left(\alpha^{\prime}\lambda^{2}+2\right)+4\alpha^{\prime2}\lambda^{4}\text{csch}^{2}\left(\lambda x\right)+2\left(\alpha^{\prime}\lambda^{2}+1\right)\cosh\left(2\lambda x\right)-2\right)}{\left(\alpha^{\prime}\lambda^{2}+\cosh\left(2\lambda x\right)-1\right)^{2}}.
\end{eqnarray}

\noindent From this solution, we can obtain the scale factor:

\begin{equation}
ds^{2}=-a_{out}\left(x\right)^{2}dt^{2}+dx^{2}.\label{eq:outside solution}
\end{equation}

\noindent with

\begin{eqnarray}
a_{out}\left(x\right) & = & C_{out}\exp2\sqrt{2}\left[\frac{-2\alpha^{\prime}\lambda^{2}-3}{3\sqrt{\alpha^{\prime}}\lambda\left(\alpha^{\prime}\lambda^{2}+2\right)}i\mathbb{F}\left(ix\lambda\left|\frac{2}{\alpha^{\prime}\lambda^{2}}\right.\right)+\frac{\sqrt{\alpha^{\prime}}\lambda\left(2\alpha^{\prime}\lambda^{2}-5\right)}{3\left(\alpha^{\prime2}\lambda^{4}-4\right)}i\mathbb{E}\left(ix\lambda\left|\frac{2}{\alpha^{\prime}\lambda^{2}}\right.\right)+\right.\nonumber \\
 &  & \left.\frac{\sinh\left(2\lambda x\right)\left(\alpha^{\prime}\lambda^{2}\left(\alpha^{\prime}\lambda^{2}-5\right)+\left(2\alpha^{\prime}\lambda^{2}-5\right)\cosh\left(2\lambda x\right)+5\right)}{3\left(\alpha^{\prime2}\lambda^{4}-4\right)\left(\alpha^{\prime}\lambda^{2}+\cosh\left(2\lambda x\right)-1\right)^{3/2}}\right],
\end{eqnarray}

\noindent where $\mathbb{F}\left(\phi|m\right)$ and $\mathbb{E}\left(\phi|m\right)$
are elliptic integrals of the first and second kinds, and $C_{out}$
is an integral constant. The corresponding Kretschmann scalar is given
by:

\begin{eqnarray}
 &  & R_{\mu\nu\rho\sigma}R^{\mu\nu\rho\sigma}=\frac{1}{2}R_{\mu\nu}R^{\mu\nu}=R^{2}=4\left(\frac{\ddot{a}}{a}\right)^{2}\nonumber \\
 & = & \frac{16\lambda^{2}}{\left(\alpha\lambda^{2}+2\right)^{2}\left(\alpha\lambda^{2}+\cosh\left(2\lambda x\right)-1\right)^{5}}\left[8\alpha^{\prime4}\lambda^{8}+16\alpha^{\prime3}\lambda^{6}\sinh^{2}\left(\lambda x\right)\left(\cosh(2\lambda x)+4\right)\right.\nonumber \\
 &  & +4\alpha^{\prime2}\lambda^{4}\sinh^{4}\left(\lambda x\right)\left(16\cosh\left(2\lambda x\right)+\cosh\left(4\lambda x\right)+41\right)-\sqrt{2}\left(\alpha^{\prime}\lambda^{2}+2\right)\sinh^{3}\left(\lambda x\right)\cosh\left(\lambda x\right)\times\nonumber \\
 &  & \sqrt{\alpha^{\prime}\lambda^{2}+\cosh\left(2\lambda x\right)-1}\left(\alpha^{\prime}\lambda^{2}\left(4\alpha^{\prime}\lambda^{2}\left(7-2\alpha^{\prime}\lambda^{2}\right)-21\right)\right.\nonumber \\
 &  & \left.+\left(-6\alpha^{\prime2}\lambda^{4}+20\alpha^{\prime}\lambda^{2}-4\right)\cosh\left(2\lambda x\right)+\left(\alpha^{\prime}\lambda^{2}+1\right)\cosh\left(4\lambda x\right)+3\right)\nonumber \\
 &  & \left.+32\alpha^{\prime}\lambda^{2}\sinh^{6}\left(\lambda x\right)(\cosh\left(2\lambda x\right)+4)+32\sinh^{8}\left(\lambda x\right)\right].\nonumber \\
\end{eqnarray}

\begin{figure}[H]
\begin{centering}
\includegraphics[scale=0.4]{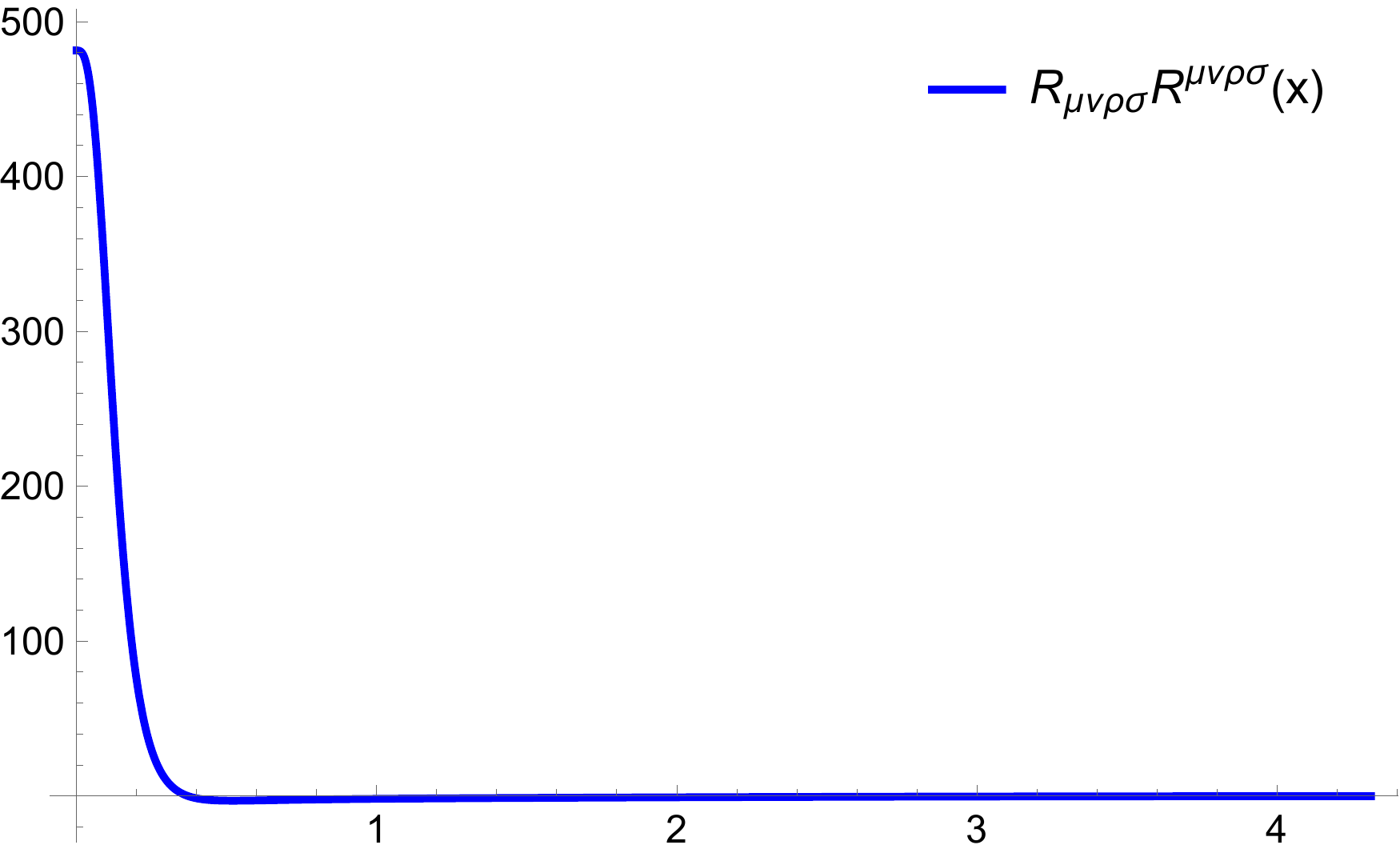}
\par\end{centering}
\centering{}\caption{\label{fig:scalar curvature outside} The figure of Kretschmann scalar,
where $\lambda=1$ and $\alpha^{\prime}=\frac{1}{16}$.}
\end{figure}

\noindent The Kretschmann scalar is displayed in Figure (\ref{fig:scalar curvature outside}).
It is demonstrates that the solution outside the event horizon ($x\geq0$)
remains regular and approaches an asymptotically flat behavior as
$x\rightarrow\infty$.

Finally, we wish establish a junction between the outside solution
(\ref{eq:outside solution}) and the inside solution (\ref{eq:inside solution})
at the event horizon. Achieving this connection implies that the event
horizon remains unaffected by the $\alpha^{\prime}$ corrections.
Specifically, since the event horizon is located at $x=0$ in both
solutions, we require $a_{in}\left(0\right)=a_{out}\left(0\right)$,
leading to the following constraint:

\begin{equation}
C_{in}=C_{out}.
\end{equation}

In summary, we have the following results in this subsection:
\begin{itemize}
\item The event horizon is not affected by the $\alpha^{\prime}$ corrections.
This result is consistent with the findings in the Schwarzschild gauge.
\item The $\alpha^{\prime}$ corrections do not introduce additional singularities
outside the event horizon.
\item The singularity of the dilaton in the perturbative solution is not
removed by the $\alpha^{\prime}$ corrections.
\end{itemize}

\section{Conclusion}

In this paper, we calculated the non-perturbative and non-singular
string black hole solutions in both the Schwarzschild gauge and unitary
gauge. The results obtained from these two gauges were in agreement,
and they can be transformed into each other through appropriate coordinate
transformations. Our findings highlight that the $\alpha^{\prime}$
corrections in string theory are capable of eliminating the curvature
singularities of two-dimensional black holes. Furthermore, in the
perturbative limit $\alpha^{\prime}\rightarrow0$, our solution matches
the divergent solutions derived from the tree-level string effective
action. The implications of our research are as follows:
\begin{itemize}
\item The curvature singularity of the two-dimensional black hole can be
removed.
\item The event horizon remains unaffected by the $\alpha^{\prime}$ corrections.
\item The $\alpha^{\prime}$ corrections do not introduce any additional
singularities to the two-dimensional black hole.
\end{itemize}
In the following works, it is worthwhile to explore the following
issues:
\begin{itemize}
\item How to remove the curvature singularities of more general black holes
is a significant problem, such as spherically symmetric black holes.
This black hole solutions are more relevant to our physical world.
However, the metric of a spherically symmetric black hole depends
on two coordinates $r$ and $\theta$, which breaks the $O\left(d,d\right)$
symmetry. The Hohm-Zwiebach action cannot be applied directly.
\item Investigating higher-dimensional black holes is of great interest.
Given that the line element of a higher-dimensional black hole is
anisotropic, the Hohm-Zwiebach action needs to incorporate the multi-trace
term, which adds complexity to the calculations. Hence, developing
a method to compute perturbative and non-perturbative solutions while
accounting for the multi-trace terms is necessary.
\item Exploring the incorporation of the Kalb-Ramond field \cite{Bernardo:2021xtr}
is also an interested direction. For instance, the BTZ black hole
solution in string theory requires a non-vanishing Kalb-Ramond field.
This implies that if we aim to eliminate the curvature singularity
in this black hole, the Hohm-Zwiebach action must include the Kalb-Ramond
field as well.
\end{itemize}
\bigskip

\vspace{5mm}

\noindent {\bf Acknowledgements}
We are deeply indebted to Xin Li, Peng Wang, Houwen Wu and Haitang Yang for many illuminating discussions and suggestions. This work is supported in part by NSFC (Grant No. 12105031), and the Postdoctoral Science Foundation of Chongqing (Grant No. cstc2021jcyj-bshX0227).

\end{document}